\documentclass[onecolumn, pra, showpacs]{revtex4}
\usepackage{graphicx}
\usepackage{dcolumn}
\usepackage{bm}
\usepackage{amsmath}

\begin{document}

\title{The rotational invariants constructed by the products\\
 of three spherical harmonic polynomials}

\author{Zhong-Qi Ma\footnote{Electronic
address: mazq@ihep.ac.cn}}
\affiliation{Institute of High Energy Physics, Chinese Academy of Sciences, Beijing 100049, China}
\author{Zong-Chao Yan\footnote{Electronic
address: zyan@unb.ca}}
\affiliation{Department of Physics, University of New Brunswick,
Fredericton, New Brunswick, Canada E3B 5A3}

\begin{abstract}
The rotational invariants constructed by the products
of three spherical harmonic polynomials are expressed generally
as homogeneous polynomials with respect to the three coordinate vectors, where the coefficients are calculated explicitly in this paper.
\end{abstract}


%02.30.Gp, Special functions
%03.65.Fd, Algebraic method
%02.20.Qs, General properties, Structure, and Representation of %          Lie groups

\date{\today}
\maketitle

\section{Introduction}

Weyl (see p.53 of \cite{Weyl}) established a theorem on the important structure for rotational invariants: Every even invariant depending on $n$ vectors ${\bf r}_{1}$, ${\bf r}_{2}$, $\ldots$, ${\bf r}_{n}$ in the three-dimensional space ${\cal R}_{3}$ is expressible in terms of the $n^{2}$ scalar products ${\bf r}_{a}\cdot {\bf r}_{b}$. Every odd invariant is a sum of terms
\begin{equation}
\left[\left({\bf r}_{a}\times {\bf r}_{b}\right)\cdot {\bf r}_{c}\right]I({\bf r}_{1},{\bf r}_{2},\ldots,{\bf r}_{n}),
\end{equation}

\noindent where ${\bf r}_{a}$, ${\bf r}_{b}$, ${\bf r}_{c}$
are selected from ${\bf r}_{1},{\bf r}_{2},\ldots,{\bf r}_{n}$,
and $I({\bf r}_{1},{\bf r}_{2},\ldots,{\bf r}_{n})$ is an even invariant. Due to the property of the rotational group SO(3), the expression for the invariant $I({\bf r}_{1},{\bf r}_{2},\ldots,{\bf r}_{n})$ is not unique except for $n\leq 3$ because it depends upon the coupled orders of $n$ vectors ${\bf r}_{a}$. In their famous Encyclopedia of Mathematics on Angular Momentum in Quantum Physics \cite{Bie}, Biedenharn and Louck studied the most important case $n=3$ of the general theorem in some detail (\S 6.17 of \cite{Bie}), and defined the even invariant (see (6.153) of \cite{Bie}) as
\begin{equation}\begin{array}{l}
I_{j,k,\ell}({\bf r}_{1},{\bf r}_{2},{\bf r}_{3})=
\sqrt{\displaystyle \frac{(4\pi)^{3}}{(2j+1)(2k+1)(2\ell+1)}}\\[3mm]
~~~\times \displaystyle \sum_{\mu,\nu,\rho}\left(
\begin{array}{ccc} j&k&\ell\\ \mu & \nu & \rho\end{array}\right){\cal Y}^{j}_{\mu}({\bf r}_{1})
{\cal Y}^{k}_{\nu}({\bf r}_{2}){\cal Y}^{\ell}_{\rho}({\bf r}_{3})\\[3mm]
~~~=\displaystyle \sum_{(\alpha)}A_{(\alpha)}\prod_{a\leq b}
\left({\bf r}_{a}\cdot {\bf r}_{b}\right)^{\alpha_{ab}},
\end{array} \label{inv} \end{equation}

\noindent where $\left(
\begin{array}{ccc} j&k&\ell\\ \mu & \nu & \rho \end{array}\right)$ is the Wigner 3-$j$ symbol, $\rho$ has to be equal to $-\mu-\nu$ owing to the property of the 3-$j$ symbol, ${\cal Y}^{j}_{\mu}({\bf r})$ denotes the spherical harmonic polynomial, and $\sum \alpha_{ab}=(j+k+\ell)/2$. The odd invariant $I_{j,k,\ell}$ is proportional to an even invariant multiplied by a factor $({\bf r}_{1}\times {\bf r}_{2})\cdot{\bf r}_{3}$. However, Biedenharn and Louck pointed out in their book (p.308 of \cite{Bie}) that: ``Unfortunately, the expression for the general coefficients (6.157) has not been given in the literature, and one has had to work out these invariant polynomials from the definition, Eq.~(6.153)", where the coefficients (6.157) mean $A_{(\alpha)}$.

As a matter of fact, there are only $(3n-3)$ independent rotational invariants constructed from $n$ coordinate vectors ${\bf r}_{a}$. The $n(n+1)/2$ different scalar products ${\bf r}_{a}\cdot {\bf r}_{b}$ are obviously too many to be independent for $n\geq 4$. The independent and complete set of rotational invariants constructed from $n$ coordinate vectors ${\bf r}_{a}$ is
\cite{gdm}:
\begin{equation}
{\bf r}_{1}\cdot {\bf r}_{a},\qquad {\bf r}_{2}\cdot {\bf r}_{b},\qquad \left({\bf r}_{1}\times {\bf r}_{2}\right)\cdot {\bf r}_{c},\label{set} \end{equation}

\noindent where $1\leq a \leq n$, $2\leq b \leq n$, and $3\leq c \leq n$. When $n=3$, the set (\ref{set}) becomes
\begin{eqnarray}
&&\eta_{1}={\bf r}_{2}\cdot {\bf r}_{3},~~~\eta_{2}={\bf r}_{3}\cdot {\bf r}_{1},~~~\eta_{3}={\bf r}_{1}\cdot {\bf r}_{2},\nonumber\\[2mm]
&&\xi_{1}={\bf r}_{1}\cdot {\bf r}_{1},
~~~\xi_{2}={\bf r}_{2}\cdot {\bf r}_{2},~~~
\zeta=\left({\bf r}_{1}\times {\bf r}_{2}\right)
\cdot {\bf r}_{3}, \label{3inv}\end{eqnarray}

\noindent where $\zeta$ has odd parity and the remaining have even parity. Due to the identity
\begin{eqnarray}
\zeta^{2}=\xi_{1}\xi_{2}\xi_{3}
-\xi_{1}\eta_{1}^{2}-\xi_{2}\eta_{2}^{2}-\xi_{3}\eta_{3}^{2}
+2\eta_{1}\eta_{2}\eta_{3},
\end{eqnarray}

\noindent with $\xi_{3}={\bf r}_{3}\cdot {\bf r}_{3}$, the invariant $\zeta$ may be replaced by $\xi_{3}$ in (\ref{inv}) for the even invariant $I_{j,k,\ell}({\bf r}_{1},{\bf r}_{2},{\bf r}_{3})$.

The rotational invariant $I_{j,k,\ell}$ has widespread applications in many branches of
physics, such as in atomic physics~\cite{atom1,atom2}, in nuclear reaction~\cite{Bie}, in condensed matter physics~\cite{condense1}, in cosmology~\cite{cosmic1}, and in astronomy and astrophysics~\cite{astro1,astro2,astro3}. Thus, it is desirable to find an explicit expression for $I_{j,k,\ell}$ in terms of the rotationally invariant variables $\eta_i$ and $\xi_i$.

At first sight, the calculation for $I_{j,k,\ell}$ may be simplified by choosing a special rotation because the spherical harmonic polynomial ${\cal Y}^{j}_{\mu}({\bf r})$ may become much simpler when ${\bf r}$ points to some special direction, say along the $z$-axis. Making a rotation such that ${\bf r}_{1}$ points to the $z$-axis and ${\bf r}_{2}$ is in the $xz$ plane with
non-negative $x$-component, one is able to express the invariant $I_{j,k,\ell}$ in terms of the product of special functions (see Appendix). With this idea, Harris~\cite{Har} calculated the even invariants $I_{j,k,\ell}$. His expression for the even $I_{j,k,\ell}$ is written in terms of the angles among the three vectors ${\bf r}_{1}$, ${\bf r}_{2}$ and ${\bf r}_{3}$, but not in terms of the scalar products ${\bf r}_{a}\cdot {\bf r}_{b}$ directly. Furthermore, his result contains coefficients $C_{j}$ which are not given explicitly (see (20) in \cite{Har}). Finally, Harris did not calculate the case of the odd invariants $I_{j,k,\ell}$.
It should be noted that the Harris approach of using spherical variables does not provide the coefficients of the invariant polynomials explicitly so that his method does not improve the way of working out these invariant polynomials from the definition,
as pointed out by Biedenharn and Louck (p.308 of \cite{Bie}).

The purpose of this paper is to present an independent calculation of the coefficients $A_{(\alpha)}$ for both even and odd invariants $I_{j,k,\ell}$ using group theoretical method. The general properties of these invariants are listed in Sec.~\ref{sec:general}. The coefficients $A_{(\alpha)}$ for even and odd invariants are calculated in Sec.~\ref{sec:even} and Sec.~\ref{sec:odd}, respectively. The conclusions are given in Sec.~\ref{sec:con}. In Appendix the invariants are expressed in terms of the angles and lengths among three coordinate vectors.

\section{General Properties of the invariants}
\label{sec:general}

For any given three non-negative integers $j$, $k$, and $\ell$,
satisfying the ``triangle rule":
\begin{equation}
|j-k|\leq \ell\leq j+k, \label{tri}
\end{equation}

\noindent the rotational invariant $I_{j,k,\ell}({\bf r}_{1},{\bf r}_{2},{\bf r}_{3})$ constructed from the products of three spherical harmonic polynomials is defined in (\ref{inv}). The invariant $I_{j,k,\ell}({\bf r}_{1},{\bf r}_{2},{\bf r}_{3})$ has the following properties.

a) $I_{j,k,\ell}({\bf r}_{1},{\bf r}_{2},{\bf r}_{3})$ is a homogeneous polynomial of orders $j$, $k$, and $\ell$ with respect to the coordinate vectors ${\bf r}_{1}$, ${\bf r}_{2}$, and ${\bf r}_{3}$, respectively.

b) The parity of $I_{j,k,\ell}({\bf r}_{1},{\bf r}_{2},{\bf r}_{3})$ is $(-1)^{j+k+\ell}$.

c) Due to the symmetry of the Wigner 3-$j$ symbol
\begin{eqnarray}
&&(-1)^{j+k+\ell}\left(\begin{array}{ccc}j & k & \ell\\ \mu & \nu &\rho \end{array} \right)=\left(\begin{array}{ccc}k & j & \ell\\ \nu & \mu & \rho \end{array} \right) \nonumber\\[2mm]
&&~~~=\left(\begin{array}{ccc}j & \ell & k\\ \mu & \rho & \nu \end{array} \right)=\left(\begin{array}{ccc}j &
k & \ell\\ -\mu & -\nu & -\rho \end{array} \right),
\nonumber \end{eqnarray}

\noindent we have
\begin{eqnarray}
&&(-1)^{j+k+\ell}I_{j,k,\ell}({\bf r}_{1},{\bf r}_{2},{\bf r}_{3})=I_{k,j,\ell}({\bf r}_{2},{\bf r}_{1},{\bf r}_{3})\nonumber \\[2mm]
&&~~~=I_{j,\ell,k}({\bf r}_{1},{\bf r}_{3},{\bf r}_{2}). \end{eqnarray}

\noindent Thus, we only need to consider $I_{j,k,\ell}$ with $j\leq k \leq \ell$. For the sake of convenience, we write an even invariant as $I_{j,k,j+k-2n}$ and an odd invariant as $I_{j+1,k+1,j+k-2n+1}$ where $0\leq 2n\leq j\leq k$.

d) $I_{j,k,j+k-2n}$ is real and $I_{j+1,k+1,j+k-2n+1}$ is pure imaginary because
\begin{eqnarray}
&&\left[\displaystyle \sum_{\mu\nu\rho}
\left(\begin{array}{ccc}j & k & \ell\\ \mu & \nu & \rho \end{array}
\right) {\cal Y}^{j}_{\mu}({\bf r}_{1}){\cal Y}^{k}_{\nu}({\bf
r}_{2}){\cal Y}^{\ell}_{\rho}({\bf r}_{3})\right]^{*}\nonumber \\[2mm]
&&~~~=\displaystyle \sum_{\mu\nu\rho}
\left(\begin{array}{ccc}j & k & \ell\\ \mu & \nu & \rho \end{array}
\right) {\cal Y}^{j}_{-\mu}({\bf r}_{1}){\cal Y}^{k}_{-\nu}({\bf
r}_{2}){\cal Y}^{\ell}_{-\rho}({\bf r}_{3}).\nonumber \end{eqnarray}

e) $I_{j,k,\ell}({\bf r}_{1},{\bf r}_{2},{\bf r}_{3})$ satisfies the three Laplace's equations with respect to ${\bf r}_{1}$, ${\bf r}_{2}$, and ${\bf r}_{3}$, respectively,
\begin{equation}
\triangle_{1}I_{j,k,\ell}=\triangle_{2}I_{j,k,\ell}
=\triangle_{3}I_{j,k,\ell}=0. \label{Lap}
\end{equation}

f) From group theory on SO(3) \cite{Bie}, the decomposition of
the direct product of three irreducible representations of SO(3), $D^{j}(R)\times D^{k}(R)\times D^{\ell}(R)$, where $j$, $k$, and $\ell$ satisfy the triangle rule (\ref{tri}), contains one and only one identity representation $D^{0}(R)$. Thus, a homogeneous polynomial of orders $j$, $k$, and $\ell$ with respect to the coordinate vectors ${\bf r}_{1}$, ${\bf r}_{2}$, and ${\bf r}_{3}$, respectively, which satisfies the Laplace's equations (\ref{Lap}), does exist and is unique up to a constant factor.

From the explicit formula for the spherical harmonic polynomial \cite{Bie}
\begin{eqnarray}
&&{\cal Y}^{k}_{\pm 1}({\bf r})=\mp \sqrt{\displaystyle \frac{
k(k+1)(2k+1)}{16\pi}}\left\{(x\pm
{\rm i}y)z^{k-1}+\ldots\right\},\nonumber \\[3mm]
&&{\cal Y}^{k}_{0}({\bf r})=\sqrt{\displaystyle \frac{
(2k+1)}{4\pi}}\left\{z^{k}+\ldots\right\},\label{func}
\end{eqnarray}

\noindent where $x$, $y$, and $z$ are the three components of ${\bf r}$. Among all spherical harmonic polynomials ${\cal Y}^{k}_{\nu}({\bf r})$ with $k$ given, the term $z^{k}$ appears only in ${\cal Y}^{k}_{0}({\bf r})$ and the terms $xz^{k-1}$ and $yz^{k-1}$ appear only in ${\cal Y}^{k}_{\pm 1}({\bf r})$. Therefore, $z_{1}^{j}z_{2}^{k}z_{3}^{j+k-2n}$ is contained only once in the homogeneous polynomial $I_{j,k,j+k-2n}({\bf r}_{1},{\bf r}_{2},{\bf r}_{3})$, at ${\cal Y}^{j}_{0}({\bf r}_{1}){\cal Y}^{k}_{0}({\bf r}_{2}){\cal Y}^{j+k-2n}_{0}({\bf r}_{3})$, with the coefficient
\begin{equation}
\left(\begin{array}{ccc} j&k&j+k-2n\\
0 & 0 & 0 \end{array}\right).
\end{equation}

\noindent Thus the even invariant defined in (\ref{inv}) can be rewritten more explicitly in the form
\begin{eqnarray}
&&I_{j,k,j+k-2n}({\bf r}_{1},{\bf r}_{2},{\bf r}_{3})
=\left(\begin{array}{ccc} j&k&j+k-2n\\
0 & 0 & 0 \end{array}\right)\nonumber \\
&&~~~\times P_{j,k,j+k-2n}^{-1}\displaystyle
\sum_{a=0}^{[j/2]}\sum_{b=0}^{j-2a}\sum_{c={\rm max}\{0,
n-a-b\}}^{[(k-b)/2]}A_{abc}\nonumber \\
&&~~~\times\xi_{1}^{a}\xi_{2}^{c} \xi_{3}^{a+b+c-n}
\eta_{1}^{k-2c-b}\eta_{2}^{j-2a-b}\eta_{3}^{b},\nonumber \\[3mm]
&&P_{j,k,j+k-2n}=\displaystyle \sum_{a}\sum_{b}\sum_{c} A_{abc}, \label{Even} \end{eqnarray}

\noindent where $[m]$ denotes the largest integer equal to or less than the non-negative real number $m$. Similarly, $x_{1}y_{2}z_{1}^{j}z_{2}^{k}z_{3}^{j+k-2n}$ is contained only twice in the homogeneous polynomial $I_{j+1,k+1,j+k-2n+1}({\bf r}_{1},{\bf r}_{2},{\bf r}_{3})$, at ${\cal Y}^{j+1}_{\pm 1}({\bf r}_{1}){\cal Y}^{k+1}_{\mp 1}({\bf r}_{2}){\cal Y}^{j+k-2n+1}_{0}({\bf r}_{3})$, with the coefficient
\begin{equation}
\displaystyle \frac{{\rm i}}{2}\sqrt{\displaystyle \frac{(j+2)!(k+2)!}{j!k!}}\left(\begin{array}{ccc} j+1&k+1&j+k-2n+1\\
1 & -1 & 0 \end{array}\right).
\end{equation}

\noindent Thus the odd invariant defined in (\ref{inv}) can be rewritten more explicitly in the form
\begin{eqnarray}
&&I_{j+1,k+1,j+k-2n+1}({\bf r}_{1},{\bf r}_{2},{\bf r}_{3})
=\displaystyle \frac{{\rm i}\zeta}{2}\sqrt{\displaystyle \frac{(j+2)!(k+2)!}{j!k!}}\nonumber\\[2mm]
&&~~~\times\left(\begin{array}{ccc}
j+1&k+1&j+k-2n+1\\1&-1&0\end{array}\right)\nonumber\\[2mm]
&&~~~\times Q_{j,k,j+k-2n}^{-1}\displaystyle
\sum_{a=0}^{[j/2]}\sum_{b=0}^{j-2a}\sum_{c={\rm max}\{0,
n-a-b\}}^{[(k-b)/2]}B_{abc}\nonumber \\
&&~~~\times\xi_{1}^{a}\xi_{2}^{c} \xi_{3}^{a+b+c-n}
\eta_{1}^{k-2c-b}\eta_{2}^{j-2a-b}\eta_{3}^{b},\nonumber \\[3mm]
&&Q_{j,k,j+k-2n}=\displaystyle \sum_{a}\sum_{b}\sum_{c} B_{abc}. \label{Odd}\end{eqnarray}

\section{Even Invariants}
\label{sec:even}

The coefficients $A_{abc}$ in (\ref{Even}) are determined from the conditions (\ref{Lap}), leading directly to the following recursive relations:
\begin{eqnarray}
&&2a(2j-2a+1)A_{abc}\nonumber \\[2mm]
&&+(j-2a-b+2)(j-2a-b+1)A_{(a-1)bc}\nonumber\\[2mm]
&&+(b+2)(b+1)A_{(a-1)(b+2)(c-1)} \nonumber \\[2mm]
&&+2(b+1)(j-2a-b+1)A_{(a-1)(b+1)c}=0,\label{eq1}
\end{eqnarray}
\begin{eqnarray}
&&2c(2k-2c+1)A_{abc}\nonumber\\[2mm]
&&+(k-2c-b+2)(k-2c-b+1) A_{ab(c-1)}\nonumber\\[2mm]
&&+(b+2)(b+1)A_{(a-1)(b+2)(c-1)}\nonumber\\[2mm]
&&+2(b+1)(k-2c-b+1)A_{a(b+1)(c-1)}=0,\label{eq2}
\end{eqnarray}
\begin{eqnarray}
&&2(a+b+c-n)(2k+2j-2a-2b-2c-2n+1)A_{abc}\nonumber\\[2mm]
&&+(k-2c-b+2)(k-2c-b+1)
A_{ab(c-1)} \nonumber\\[2mm]
&&+(j-2a-b+2)(j-2a-b+1)A_{(a-1)bc}\nonumber\\[2mm]
&&+2(k-2c-b+1)(j-2a-b+1)A_{a(b-1)c}=0.\label{eq3}
\end{eqnarray}

\noindent Due to the property f) in Sec.~\ref{sec:general}, the solutions for $A_{abc}$ exist uniquely up to a common numerical factor, which can be determined for convenience by
\begin{eqnarray}
&&A_{00n}=\displaystyle \frac{(k-n)!(2j-1)!!}{(k-2n)!(2j-2n-1)!!}
\nonumber\\[2mm]
&&~~~ \times \displaystyle \prod_{m=1}^{\lambda}2m(2j+2k-4n-2m+1),\\[2mm]
&&\lambda=[(j+k)/2]-n.
\end{eqnarray}

\noindent The constant $\lambda$ is chosen to be larger enough so that the coefficients $A_{abc}$ do not contain an unnecessary denominator. On the other hand, if $\lambda$ is too large, it may make $A_{abc}$ to be more complex due to the existence of a common factor. The choice of $\lambda$, however, does not effect the final results.

The coefficients $A_{abc}$ can be calculated one by one by mathematical induction from the recursive relations (\ref{eq1}-\ref{eq3}), where the calculation order is critical. The key point is that at each step of calculation using one of (\ref{eq1}-\ref{eq3}), only one {\it unknown} coefficient is solved from three remaining {\it known} coefficients. We choose the calculation order as follows and list the calculation results.

First, we calculate $A_{0b(n-b)}$ from (\ref{eq2}), and $A_{ab(n-a-b)}$ from (\ref{eq1}) by mathematical induction:
\begin{eqnarray}
&&A_{ab(n-a-b)}=G_{a,b} \displaystyle \frac{(j-n)!(k-n)!} {(j-2a-b)!(k-2n+2a+b)!}\nonumber\\[2mm]
&&~~~ \times \displaystyle \prod_{m=1}^{\lambda}2m(2j+2k-4n-2m+1), \label{e-1}
\end{eqnarray}

\noindent where $0\leq a \leq n$, $0\leq b \leq n-a$, and
\begin{eqnarray}
&&G_{a,b}=\displaystyle \sum_{r=\max\{0,2a+b-n\}}^{a}
\displaystyle \frac{(-1)^{a+b+r}n!} {2^{a-r}r!(a-r)!b!}\nonumber\\[2mm]
&& \times \displaystyle \frac{(j-2a-b+r)!(k-2n+2a+b)!} {(n-2a-b+r)!(j-n)!(k-2n+2a+b-r)!}\nonumber\\[2mm]
&& \times \displaystyle \frac{(2j-2a-1)!!(2k-2n+4a+2b-2r-1)!!} {(2j-2n-1)!!(2k-2n-1)!!}. \label{e-g}
\end{eqnarray}

\noindent Evidently, $G_{a,b}=0$ if $a<0$, or $b<0$, or $a+b>n$. The function $G_{a,b}$ will play an essential role for
later calculations.

Second, we calculate $A_{ab(n-a-b+c)}$ from (\ref{eq3}) by mathematical induction:
\begin{eqnarray}
&&A_{ab(n-a-b+c)}=\displaystyle \sum_{s=0}^{b}
\sum_{r=\max\{s,c-a\}}^{c}G_{a-c+r,b-s}\nonumber\\[3mm]
&& \times \displaystyle \frac{(-1)^{c}2^{s}(c!)(j-n)!(k-n)!} {(c-r)!(r-s)!s!(j-2a-b)!(k-2n+2a+b-2c)!}\nonumber\\[3mm]
&& \times \displaystyle \prod_{m=c+1}^{\lambda}2m(2j+2k-4n-2m+1), \label{e-2}
\end{eqnarray}

\noindent where $0\leq a \leq n$, $0\leq b \leq n-a$, and $0\leq c\leq [(k+b)/2]+a-n$. For the case of $c>a$, terms with $c-a>r\geq s$, which may occur in the sum over $r$ in (\ref{e-2}), vanish because $G_{a,b}=0$ if $a<0$. It is for this reason that the lower bound of summation over $r$ in $(\ref{e-2})$ becomes $\max\{s,c-a\}$. Similar cases occur in the following formulas.

Third, we calculate $A_{a(n-a+b)c}$ from (\ref{eq3}) by mathematical induction:
\begin{eqnarray}
&&A_{a(n-a+b)c}=\displaystyle \sum_{s=\max\{0,b-a\}}^{\min\{b+c,n-a+b\}} \sum_{r=\max\{s,b+c-a\}}^{\min\{s+c,b+c\}}
\nonumber\\[3mm]
&&~~~ G_{a-b-c+r,n-a+b-s}\displaystyle \frac{(-1)^{b+c}2^{s}(b+c)!}{(r-s)!s!(b+c-r)!}\nonumber\\[3mm]
&&~~~ \times \displaystyle \frac{(j-n)!(k-n)!} {(j-n-a-b)!(k-n+a-b-2c)!}\nonumber\\[3mm]
&&~~~ \times \displaystyle \prod_{m=b+c+1}^{\lambda}2m(2j+2k-4n-2m+1), \label{e-3}
\end{eqnarray}

\noindent where $0\leq a \leq n$,
$0 \leq b \leq j-n-a$, and $0\leq c \leq [(k-n+a-b)/2]$.
 For the case of $b+c>a$, terms with $s\leq r< b+c-a$, which may occur in the sum over $r$ in (\ref{e-3}), vanish because $G_{a,b}=0$ if $a<0$. When $s>b$, terms with $b+c<r\leq s+c$ vanish due to the existence of the factor $(b+c-r)!$ at the denominator. For the case of $b>a$, terms with $0\leq s<b-a$, which may occur in the sum over $s$ in (\ref{e-3}), vanish because $G_{a,b}=0$ if $b>n$. Finally, for the case of $c>n-a$, terms with $n-a+b< s\leq b+c$, which may occur in the sum over $s$ in (\ref{e-3}), vanish because $G_{a,b}=0$ if $b<0$.

Finally, we calculate $A_{(n+a)bc}$ from (\ref{eq3}) by mathematical induction:
\begin{eqnarray}
&&A_{(n+a)bc}=\displaystyle \sum_{s=\max\{0,b-n\}}^{b}
\sum_{r=\max\{s,b+c-n\}}^{s+c}\nonumber\\[3mm]
&&~~~ G_{n-b-c+r,b-s}\displaystyle \frac{(-1)^{a+b+c}2^{s}(a+b+c)!}{(r-s)!s!(a+b+c-r)!}\nonumber\\[3mm]
&&~~~ \times \displaystyle \frac{(j-n)!(k-n)!} {(j-2n-2a-b)!(k-2c-b)!}\nonumber\\[3mm]
&&~~~ \times \displaystyle \prod_{m=a+b+c+1}^{\lambda}2m(2j+2k-4n-2m+1), \label{e-4}
\end{eqnarray}

\noindent where $0\leq a \leq [j/2]-n$, $0\leq b \leq j-2n-2a$, and $0\leq c \leq [(k-b)/2]$. For the case of $b+c>n$, terms with $s\leq r< b+c-n$, which may occur in the sum over $r$ in (\ref{e-4}), vanish because $G_{a,b}=0$ if $a<0$. For the case of $b>n$,  terms with $0\leq s< b-n$, which may occur in the sum over $s$ in (\ref{e-4}), vanish because $G_{a,b}=0$ if $b>n$.

In the following we list some special even invariants $I_{j,k,j+k-2n}$ for reference.
$$\begin{array}{rl}
I_{0,0,0}&=1,\qquad I_{0,1,1}=\displaystyle
\frac{-1}{\sqrt{3}}~\eta_{1}, \\[2mm]
I_{0,2,2}&=\displaystyle
\frac{1}{\sqrt{5}}\left\{ \displaystyle
\frac{1}{2}\left[3\eta_{1}^{2}-\xi_{2}\xi_{3}\right]\right\},
 \\[2mm]
I_{0,3,3}&=\displaystyle
\frac{-1}{\sqrt{7}}\left\{  \displaystyle
\frac{1}{2}\left[5\eta_{1}^{3}-3\xi_{2}\xi_{3}\eta_{1}
\right]\right\}, \\[2mm]
I_{0,4,4}&=\displaystyle
\frac{1}{3}\left\{ \displaystyle
\frac{1}{8}\left[35\eta_{1}^{4}-30\xi_{2}\xi_{3}\eta_{1}^{2}
+3\xi_{2}^{2}\xi_{3}^{2}\right]\right\}, \\[2mm]
I_{0,5,5}&=\displaystyle
\frac{-1}{\sqrt{11}}\left\{ \displaystyle
\frac{1}{8}\left[63\eta_{1}^{5}-70\xi_{2}\xi_{3}\eta_{1}^{3}
+15\xi_{2}^{2}\xi_{3}^{2}\eta_{1}\right]\right\},\\[2mm]
I_{0,6,6}&=\displaystyle
\frac{1}{\sqrt{13}}\left\{ \displaystyle
\frac{1}{16}\left[231\eta_{1}^{6}-315\xi_{2}\xi_{3}\eta_{1}^{4}
\right.\right. \\[2mm]
&\left.\left.~~~~~+105\xi_{2}^{2}\xi_{3}^{2}\eta_{1}^{2}
-5\xi_{2}^{3}\xi_{3}^{3}
\right]\right\},\\[2mm]
I_{112}&=\sqrt{\displaystyle
\frac{2}{15}}\left\{ \displaystyle
\frac{1}{2}\left[3\eta_{1}\eta_{2}-\xi_{3}\eta_{3}\right]\right\},\\[3mm]
I_{123}&=-\sqrt{\displaystyle
\frac{3}{35}}\left\{ \displaystyle
\frac{1}{2}\left[5\eta_{1}^{2}\eta_{2}
-\xi_{2}\xi_{3}\eta_{2}-2\xi_{3}\eta_{1}\eta_{3}\right]\right\},\\[3mm]
I_{134}&=\displaystyle
\frac{2}{3}\sqrt{\displaystyle
\frac{1}{7}}\left\{\displaystyle
\frac{1}{8}\left[35\eta_{1}^{3}\eta_{2}
-15\xi_{2}\xi_{3}\eta_{1}\eta_{2}-15\xi_{3}
\eta_{1}^{2}\eta_{3}\right.\right.\\[3mm]
&\left.\left.~~~+3\xi_{2}\xi_{3}^{2}\eta_{3}\right]\right\},\\[3mm]
I_{145}&=-\displaystyle
\frac{1}{3}\sqrt{\displaystyle
\frac{5}{11}}\left\{\displaystyle
\frac{1}{8}\left[63\eta_{1}^{4}\eta_{2}
-42\xi_{2}\xi_{3}\eta_{1}^{2}\eta_{2}\right.\right.\\[3mm]
&\left.\left.~~~+3\xi_{2}^{2}\xi_{3}^{2}\eta_{2}
-28\xi_{3}\eta_{1}^{3}\eta_{3} +12\xi_{2}\xi_{3}^{2}\eta_{1}\eta_{3}\right] \right\},\\[3mm]
I_{156}&=\sqrt{\displaystyle
\frac{6}{143}}\left\{ \displaystyle
\frac{1}{16}\left[231\eta_{1}^{5}\eta_{2}
-210\xi_{2}\xi_{3}\eta_{1}^{3}\eta_{2}
+35\xi_{2}^{2}\xi_{3}^{2}\eta_{1}\eta_{2}\right.\right.\\[3mm]
&\left.\left.~~~-105\xi_{3}\eta_{1}^{4}\eta_{3}
+70\xi_{2}\xi_{3}^{2}\eta_{1}^{2}\eta_{3}
-5\xi_{2}^{2}\xi_{3}^{3}\eta_{3}\right]\right\},\\[3mm]
I_{222}&=-\sqrt{\displaystyle
\frac{2}{35}}\left\{\displaystyle
\frac{1}{2}\left[-3\xi_{2}\eta_{2}^{2}
+9\eta_{1}\eta_{2}\eta_{3}-3
\xi_{3}\eta_{3}^{2}\right.\right.\\[2mm]
&~~~\left.\left.-3\xi_{1}\eta_{1}^{2}+2
\xi_{1}\xi_{2}\xi_{3}\right]\right\},\\[3mm]
I_{224}&=\sqrt{\displaystyle
\frac{2}{35}}\left\{ \displaystyle
\frac{1}{8}\left[35\eta_{1}^{2}\eta_{2}^{2}
-5\xi_{2}\xi_{3}\eta_{2}^{2}-20
\xi_{3}\eta_{1}\eta_{2}\eta_{3}\right.\right.\\[2mm]
&~~~\left.\left.+2\xi_{3}^{2}\eta_{3}^{2}-5
\xi_{1}\xi_{3}\eta_{1}^{2}+
\xi_{1}\xi_{2}\xi_{3}^{2}\right]\right\},\\[3mm]
I_{233}&=\displaystyle
\frac{2}{\sqrt{105}}\left\{ \displaystyle
\frac{1}{8}\left[-30\xi_{2}\eta_{1}\eta_{2}^{2}
+75\eta_{1}^{2}\eta_{2}\eta_{3}
-3\xi_{2}\xi_{3}\eta_{2}\eta_{3}\right.\right.\\[2mm]
&~~~\left.\left.-30\xi_{3}\eta_{1}\eta_{3}^{2}
-25\xi_{1}\eta_{1}^{3}
+21\xi_{1}\xi_{2}\xi_{3}\eta_{1}\right]\right\},\\[3mm]
I_{235}&=-\sqrt{\displaystyle
\frac{10}{231}}\left\{ \displaystyle
\frac{1}{8}\left[63\eta_{1}^{3}\eta_{2}^{2}
-21\xi_{2}\xi_{3}\eta_{1}\eta_{2}^{2}
-42\xi_{3}\eta_{1}^{2}\eta_{2}\eta_{3}\right.\right.\\[2mm]
&~~~\left.\left.+6\xi_{2}\xi_{3}^{2}\eta_{2}\eta_{3}
+6\xi_{3}^{2}\eta_{1}\eta_{3}^{2}
-7\xi_{1}\xi_{3}\eta_{1}^{3}
+3\xi_{1}\xi_{2}\xi_{3}^{2}\eta_{1}\right]\right\},\\[3mm]
I_{244}&=\displaystyle
\frac{-2}{3}\sqrt{\displaystyle
\frac{5}{77}}\left\{\displaystyle
\frac{1}{8}\left[-63\xi_{2}\eta_{1}^{2}\eta_{2}^{2}
+9\xi_{2}^{2}\xi_{3}\eta_{2}^{2}
+147\eta_{1}^{3}\eta_{2}\eta_{3}\right.\right.\\[2mm]
&~~~-27\xi_{2}\xi_{3}\eta_{1}\eta_{2}\eta_{3}
-63\xi_{3}\eta_{1}^{2}\eta_{3}^{2}
+9\xi_{2}\xi_{3}^{2}\eta_{3}^{2}
-49\xi_{1}\eta_{1}^{4}\\[2mm]
&\left.\left.~~~+51\xi_{1}\xi_{2}\xi_{3}\eta_{1}^{2}
-6\xi_{1}\xi_{2}^{2}\xi_{3}^{2}\right]\right\},\\[3mm]
I_{246}&=\sqrt{\displaystyle
\frac{5}{143}}\left\{\displaystyle
\frac{1}{16}\left[231\eta_{1}^{4}\eta_{2}^{2}
-126\xi_{2}\xi_{3}\eta_{1}^{2}\eta_{2}^{2}
+7\xi_{2}^{2}\xi_{3}^{2}\eta_{2}^{2}\right.\right.\\[2mm]
&~~~-168\xi_{3}\eta_{1}^{3}\eta_{2}\eta_{3}
+56\xi_{2}\xi_{3}^{2}\eta_{1}\eta_{2}\eta_{3}
+28\xi_{3}^{2}\eta_{1}^{2}\eta_{3}^{2}\\[2mm]
&\left.\left.~~~-4\xi_{2}\xi_{3}^{3}\eta_{3}^{2}
-21\xi_{1}\xi_{3}\eta_{1}^{4}
+14\xi_{1}\xi_{2}\xi_{3}^{2}\eta_{1}^{2}
-\xi_{1}\xi_{2}^{2}\xi_{3}^{3}\right]\right\},\\[3mm]
\end{array}$$
$$\begin{array}{rl}
I_{334}&=-\sqrt{\displaystyle
\frac{2}{77}}\left\{ \displaystyle
\frac{1}{8}\left[-70\xi_{2}\eta_{1}\eta_{2}^{3}
+175\eta_{1}^{2}\eta_{2}^{2}\eta_{3}
+5\xi_{2}\xi_{3}\eta_{2}^{2}\eta_{3}\right.\right.\\[2mm]
&~~~-100\xi_{3}\eta_{1}\eta_{2}\eta_{3}^{2}
+10\xi_{3}^{2}\eta_{3}^{3}-70\xi_{1}\eta_{1}^{3}\eta_{2}\\[2mm]
&~~~\left.\left.
+60\xi_{1}\xi_{2}\xi_{3}\eta_{1}\eta_{2}
+5\xi_{1}\xi_{3}\eta_{1}^{2}\eta_{3}
-7\xi_{1}\xi_{2}\xi_{3}^{2}\eta_{3}
\right]\right\},\\[3mm]
I_{336}&=\displaystyle
\frac{10}{\sqrt{3003}}\left\{ \displaystyle
\frac{1}{16}\left[231\eta_{1}^{3}\eta_{2}^{3}
-63\xi_{2}\xi_{3}\eta_{1}\eta_{2}^{3}
-189\xi_{3}\eta_{1}^{2}\eta_{2}^{2}\eta_{3}\right.\right.\\[2mm]
&~~~+21\xi_{2}\xi_{3}^{2}\eta_{2}^{2}\eta_{3}
+42\xi_{3}^{2}\eta_{1}\eta_{2}\eta_{3}^{2}-2\xi_{3}^{3}\eta_{3}^{3}
-63\xi_{1}\xi_{3}\eta_{1}^{3}\eta_{2}\\[2mm]
&~~~\left.\left.+21\xi_{1}\xi_{2}\xi_{3}^{2}\eta_{1}\eta_{2}
+21\xi_{1}\xi_{3}^{2}\eta_{1}^{2}\eta_{3}
-3\xi_{1}\xi_{2}\xi_{3}^{3}\eta_{3}
\right]\right\}.\\[3mm]

\end{array}$$

\section{Odd Invariants}
\label{sec:odd}

Substituting (\ref{Odd}) into the Laplace's equations (\ref{Lap}), we obtain the recursive relations for the coefficients $B_{abc}$

\begin{eqnarray}
&&2a(2j-2a+3)B_{abc}\nonumber \\[2mm]
&&+(j-2a-b+2)(j-2a-b+1)B_{(a-1)bc}\nonumber\\[2mm]
&&+(b+2)(b+1)B_{(a-1)(b+2)(c-1)} \nonumber \\[2mm]
&&+2(b+1)(j-2a-b+1)B_{(a-1)(b+1)c}=0,\label{eq4}
\end{eqnarray}
\begin{eqnarray}
&&2c(2k-2c+3)B_{abc}\nonumber\\[2mm]
&&+(k-2c-b+2)(k-2c-b+1) B_{ab(c-1)}\nonumber\\[2mm]
&&+(b+2)(b+1)B_{(a-1)(b+2)(c-1)}\nonumber\\[2mm]
&&+2(b+1)(k-2c-b+1)B_{a(b+1)(c-1)}=0,\label{eq5}
\end{eqnarray}
\begin{eqnarray}
&&2(a+b+c-n)(2k+2j-2a-2b-2c-2n+3)B_{abc}\nonumber\\[2mm]
&&+(k-2c-b+2)(k-2c-b+1)
B_{ab(c-1)} \nonumber\\[2mm]
&&+(j-2a-b+2)(j-2a-b+1)B_{(a-1)bc}\nonumber\\[2mm]
&&+2(k-2c-b+1)(j-2a-b+1)B_{a(b-1)c}=0.\label{eq6}
\end{eqnarray}

\noindent The only difference between the two sets of equations
(\ref{eq1}-\ref{eq3}) and (\ref{eq4}-\ref{eq6}) is that ``$(+1)$" in the first term of each equation of the first set is replaced by ``$(+3)$" in that of the second set. Thus, through the same procedure we have calculated the coefficients $B_{abc}$ listed below, which are very similar to the coefficients $A_{abc}$.
\begin{eqnarray}
&&B_{00n}=\displaystyle \frac{(k-n)!(2j+1)!!}{(k-2n)!(2j-2n+1)!!}
\nonumber\\[2mm]
&&~~~ \times \displaystyle \prod_{m=1}^{\lambda}2m(2j+2k-4n-2m+3).
\end{eqnarray}
\begin{eqnarray}
&&B_{ab(n-a-b)}=F_{a,b} \displaystyle \frac{(j-n)!(k-n)!} {(j-2a-b)!(k-2n+2a+b)!}\nonumber\\[2mm]
&&~~~ \times \displaystyle \prod_{m=1}^{\lambda}2m(2j+2k-4n-2m+3), \label{o-1}
\end{eqnarray}

\noindent where $0\leq a\leq n$, $0\leq b\leq n-a$, and
\begin{eqnarray}
&&F_{a,b}=\displaystyle \sum_{r=\max\{0,2a+b-n\}}^{a}
\displaystyle \frac{(-1)^{a+b+r}n!} {2^{a-r}r!(a-r)!b!}\nonumber\\[2mm]
&& \times \displaystyle \frac{(j-2a-b+r)!(k-2n+2a+b)!} {(n-2a-b+r)!(j-n)!(k-2n+2a+b-r)!}\nonumber\\[2mm]
&& \times \displaystyle \frac{(2j-2a+1)!!(2k-2n+4a+2b-2r+1)!!} {(2j-2n+1)!!(2k-2n+1)!!}. \label{o-f}
\end{eqnarray}

\noindent Evidently, $F_{a,b}=0$ if $a<0$, or $b<0$, or $a+b>n$. \begin{eqnarray}
&&B_{ab(n-a-b+c)}=\displaystyle \sum_{s=0}^{b}
\sum_{r=\max\{s,c-a\}}^{c}F_{a-c+r,b-s}\nonumber\\[3mm]
&& \times \displaystyle \frac{(-1)^{c}2^{s}(c!)(j-n)!(k-n)!} {(c-r)!(r-s)!s!(j-2a-b)!(k-2n+2a+b-2c)!}\nonumber\\[3mm]
&& \times \displaystyle \prod_{m=c+1}^{\lambda}2m(2j+2k-4n-2m+3), \label{o-2}
\end{eqnarray}

\noindent where $0\leq a \leq n$, $0\leq b \leq n-a$, and $0\leq c\leq [(k+b)/2]+a-n$.
\begin{eqnarray}
&&B_{a(n-a+b)c}=\displaystyle \sum_{s=\max\{0,b-a\}}^{\min\{b+c,n-a+b\}} \sum_{r=\max\{s,b+c-a\}}^{\min\{s+c,b+c\}}
\nonumber\\[3mm]
&&~~~ F_{a-b-c+r,n-a+b-s}\displaystyle \frac{(-1)^{b+c}2^{s}(b+c)!}{(r-s)!s!(b+c-r)!}\nonumber\\[3mm]
&&~~~ \times \displaystyle \frac{(j-n)!(k-n)!} {(j-n-a-b)!(k-n+a-b-2c)!}\nonumber\\[3mm]
&&~~~ \times \displaystyle \prod_{m=b+c+1}^{\lambda}2m(2j+2k-4n-2m+3), \label{o-3}
\end{eqnarray}

\noindent where $0\leq a \leq n$,
$0 \leq b \leq j-n-a$, and $0\leq c \leq [(k-n+a-b)/2]$.
\begin{eqnarray}
&&B_{(n+a)bc}=\displaystyle \sum_{s=\max\{0,b-n\}}^{b}
\sum_{r=\max\{s,b+c-n\}}^{s+c}\nonumber\\[3mm]
&&~~~ F_{n-b-c+r,b-s}\displaystyle \frac{(-1)^{a+b+c}2^{s}(a+b+c)!}{(r-s)!s!(a+b+c-r)!}\nonumber\\[3mm]
&&~~~ \times \displaystyle \frac{(j-n)!(k-n)!} {(j-2n-2a-b)!(k-2c-b)!}\nonumber\\[3mm]
&&~~~ \times \displaystyle \prod_{m=a+b+c+1}^{\lambda}2m(2j+2k-4n-2m+3), \label{o-4}
\end{eqnarray}

\noindent where $0\leq a \leq [j/2]-n$, $0\leq b \leq j-2n-2a$, and $0\leq c \leq [(k-b)/2]$.

In the following we list some special odd invariants $I_{j+1,k+1,j+k-2n+1}$ for reference.

$$\begin{array}{rl}
I_{1,1,1}&=\displaystyle \frac{{\rm i}\zeta }{\sqrt{6}},\\[2mm]
I_{1,2,2}&=-{\rm i}\zeta \sqrt{\displaystyle
\frac{3}{10}}~\eta_{1}, \\[2mm]
I_{1,3,3}&={\rm i}\zeta \sqrt{\displaystyle
\frac{3}{7}}\left\{ \displaystyle
\frac{1}{4}\left[5\eta_{1}^{2}-\xi_{2}\xi_{3}\right]\right\},
 \\[2mm]
I_{1,4,4}&=-{\rm i}\zeta \displaystyle
\frac{\sqrt{5}}{3}\left\{  \displaystyle
\frac{1}{4}\left[7\eta_{1}^{3}-3\xi_{2}\xi_{3}\eta_{1}
\right]\right\}, \\[2mm]
I_{1,5,5}&={\rm i}\zeta \sqrt{\displaystyle
\frac{15}{22}}\left\{ \displaystyle
\frac{1}{8}\left[21\eta_{1}^{4}-14\xi_{2}\xi_{3}\eta_{1}^{2}
+\xi_{2}^{2}\xi_{3}^{2}\right]\right\}, \\[2mm]
I_{1,6,6}&=-{\rm i}\zeta \sqrt{\displaystyle
\frac{21}{26}}\left\{ \displaystyle
\frac{1}{8}\left[33\eta_{1}^{5}-30\xi_{2}\xi_{3}\eta_{1}^{3}
+5\xi_{2}^{2}\xi_{3}^{2}\eta_{1}\right]\right\},\\[2mm]
I_{1,7,7}&={\rm i}\zeta \sqrt{\displaystyle
\frac{14}{15}}\left\{ \displaystyle
\frac{1}{64}\left[429\eta_{1}^{6}-495\xi_{2}\xi_{3}\eta_{1}^{4}
\right.\right. \\[2mm]
&\left.\left.~~~~~+135\xi_{2}^{2}\xi_{3}^{2}\eta_{1}^{2}
-5\xi_{2}^{3}\xi_{3}^{3}
\right]\right\},\\[2mm]
I_{2,2,3}&={\rm i}\zeta 3\sqrt{\displaystyle
\frac{2}{35}}\left\{ \displaystyle
\frac{1}{4}\left[5\eta_{1}\eta_{2}-\xi_{3}\eta_{3}\right]\right\},\\[3mm]
I_{2,3,4}&=-{\rm i}\zeta \sqrt{\displaystyle
\frac{5}{7}}\left\{ \displaystyle
\frac{1}{4}\left[7\eta_{1}^{2}\eta_{2}
-\xi_{2}\xi_{3}\eta_{2}-2\xi_{3}\eta_{1}\eta_{3}\right]\right\},\\[3mm]
I_{2,4,5}&={\rm i}\zeta \sqrt{\displaystyle
\frac{10}{11}}\left\{\displaystyle
\frac{1}{8}\left[21\eta_{1}^{3}\eta_{2}
-7\xi_{2}\xi_{3}\eta_{1}\eta_{2}-7\xi_{3}
\eta_{1}^{2}\eta_{3}\right.\right.\\[3mm]
&\left.\left.~~~+\xi_{2}\xi_{3}^{2}\eta_{3}\right]\right\},\\[3mm]
I_{2,5,6}&=-{\rm i}\zeta 3\sqrt{\displaystyle
\frac{35}{286}}\left\{\displaystyle
\frac{1}{8}\left[33\eta_{1}^{4}\eta_{2}
-18\xi_{2}\xi_{3}\eta_{1}^{2}\eta_{2}\right.\right.\\[3mm]
&\left.\left.~~~+\xi_{2}^{2}\xi_{3}^{2}\eta_{2}
-12\xi_{3}\eta_{1}^{3}\eta_{3} +4\xi_{2}\xi_{3}^{2}\eta_{1}\eta_{3}\right] \right\},\\[3mm]
I_{2,6,7}&=2\sqrt{\displaystyle
\frac{21}{65}}\left\{ \displaystyle
\frac{1}{64}\left[429\eta_{1}^{5}\eta_{2}
-330\xi_{2}\xi_{3}\eta_{1}^{3}\eta_{2}
+45\xi_{2}^{2}\xi_{3}^{2}\eta_{1}\eta_{2}\right.\right.\\[3mm]
&\left.\left.~~~-165\xi_{3}\eta_{1}^{4}\eta_{3}
+90\xi_{2}\xi_{3}^{2}\eta_{1}^{2}\eta_{3}
-5\xi_{2}^{2}\xi_{3}^{3}\eta_{3}\right]\right\},\\[3mm]
I_{3,3,3}&=-{\rm i}\zeta \sqrt{\displaystyle
\frac{6}{7}}\left\{ \displaystyle
\frac{1}{12}\left[-5\xi_{2}\eta_{2}^{2}
+25\eta_{1}\eta_{2}\eta_{3}-5
\xi_{3}\eta_{3}^{2}\right.\right.\\[2mm]
&~~~\left.\left.-5\xi_{1}\eta_{1}^{2}+2
\xi_{1}\xi_{2}\xi_{3}\right]\right\},\\[3mm]
I_{3,3,5}&={\rm i}\zeta 5\sqrt{\displaystyle
\frac{3}{77}}\left\{ \displaystyle
\frac{1}{24}\left[63\eta_{1}^{2}\eta_{2}^{2}
-7\xi_{2}\xi_{3}\eta_{2}^{2}-28
\xi_{3}\eta_{1}\eta_{2}\eta_{3}\right.\right.\\[2mm]
&~~~\left.\left.+2\xi_{3}^{2}\eta_{3}^{2}-7
\xi_{1}\xi_{3}\eta_{1}^{2}+
\xi_{1}\xi_{2}\xi_{3}^{2}\right]\right\},\\[3mm]
I_{3,4,4}&={\rm i}\zeta 3\sqrt{\displaystyle
\frac{10}{77}}\left\{ \displaystyle
\frac{1}{72}\left[-70\xi_{2}\eta_{1}\eta_{2}^{2}
+245\eta_{1}^{2}\eta_{2}\eta_{3}
\right.\right.\\[2mm]
&~~~\left.\left.-15\xi_{2}\xi_{3}\eta_{2}\eta_{3}
-70\xi_{3}\eta_{1}\eta_{3}^{2}
-49\xi_{1}\eta_{1}^{3}
+31\xi_{1}\xi_{2}\xi_{3}\eta_{1}\right]\right\},\\[3mm]
I_{3,4,6}&=-{\rm i}\zeta 5\sqrt{\displaystyle
\frac{7}{143}}\left\{ \displaystyle
\frac{1}{8}\left[33\eta_{1}^{3}\eta_{2}^{2}
-9\xi_{2}\xi_{3}\eta_{1}\eta_{2}^{2}
-18\xi_{3}\eta_{1}^{2}\eta_{2}\eta_{3}\right.\right.\\[2mm]
&~~~\left.\left.+2\xi_{2}\xi_{3}^{2}\eta_{2}\eta_{3}
+2\xi_{3}^{2}\eta_{1}\eta_{3}^{2}
-3\xi_{1}\xi_{3}\eta_{1}^{3}
+\xi_{1}\xi_{2}\xi_{3}^{2}\eta_{1}\right]\right\},\\[3mm]
I_{3,5,5}&=-{\rm i}\zeta \sqrt{\displaystyle
\frac{210}{143}}\left\{\displaystyle
\frac{1}{24}\left[-45\xi_{2}\eta_{1}^{2}\eta_{2}^{2}
+5\xi_{2}^{2}\xi_{3}\eta_{2}^{2}
+135\eta_{1}^{3}\eta_{2}\eta_{3}\right.\right.\\[2mm]
&~~~-25\xi_{2}\xi_{3}\eta_{1}\eta_{2}\eta_{3}
-45\xi_{3}\eta_{1}^{2}\eta_{3}^{2}
+5\xi_{2}\xi_{3}^{2}\eta_{3}^{2}
-27\xi_{1}\eta_{1}^{4}\\[2mm]
&\left.\left.~~~+23\xi_{1}\xi_{2}\xi_{3}\eta_{1}^{2}
-2\xi_{1}\xi_{2}^{2}\xi_{3}^{2}\right]\right\},\\[3mm]
\end{array}$$
$$\begin{array}{rl}
I_{3,5,7}&={\rm i}\zeta \sqrt{\displaystyle
\frac{210}{143}}\left\{\displaystyle
\frac{1}{64}\left[429\eta_{1}^{4}\eta_{2}^{2}
-198\xi_{2}\xi_{3}\eta_{1}^{2}\eta_{2}^{2}
+9\xi_{2}^{2}\xi_{3}^{2}\eta_{2}^{2}\right.\right.\\[2mm]
&~~~-264\xi_{3}\eta_{1}^{3}\eta_{2}\eta_{3}
+72\xi_{2}\xi_{3}^{2}\eta_{1}\eta_{2}\eta_{3}
+36\xi_{3}^{2}\eta_{1}^{2}\eta_{3}^{2}
-4\xi_{2}\xi_{3}^{3}\eta_{3}^{2}\\[2mm]
&\left.\left.~~~-33\xi_{1}\xi_{3}\eta_{1}^{4}
+18\xi_{1}\xi_{2}\xi_{3}^{2}\eta_{1}^{2}
-\xi_{1}\xi_{2}^{2}\xi_{3}^{3}\right]\right\},\\[3mm]
I_{4,4,5}&=-{\rm i}\zeta \displaystyle
\frac{15}{\sqrt{143}}\left\{ \displaystyle
\frac{1}{72}\left[-126\xi_{2}\eta_{1}\eta_{2}^{3}
+441\eta_{1}^{2}\eta_{2}^{2}\eta_{3}
-7\xi_{2}\xi_{3}\eta_{2}^{2}\eta_{3}\right.\right.\\[2mm]
&~~~-196\xi_{3}\eta_{1}\eta_{2}\eta_{3}^{2}
+14\xi_{3}^{2}\eta_{3}^{3}-126\xi_{1}\eta_{1}^{3}\eta_{2}\\[2mm]
&~~~\left.\left.
+84\xi_{1}\xi_{2}\xi_{3}\eta_{1}\eta_{2}
-7\xi_{1}\xi_{3}\eta_{1}^{2}\eta_{3}
-5\xi_{1}\xi_{2}\xi_{3}^{2}\eta_{3}
\right]\right\},\\[3mm]
I_{4,4,7}&={\rm i}\zeta \displaystyle
\frac{14}{3}\sqrt{\displaystyle
\frac{10}{143}}\left\{ \displaystyle
\frac{1}{64}\left[429\eta_{1}^{3}\eta_{2}^{3}
-99\xi_{2}\xi_{3}\eta_{1}\eta_{2}^{3}
-297\xi_{3}\eta_{1}^{2}\eta_{2}^{2}\eta_{3}\right.\right.\\[2mm]
&~~~+27\xi_{2}\xi_{3}^{2}\eta_{2}^{2}\eta_{3}
+54\xi_{3}^{2}\eta_{1}\eta_{2}\eta_{3}^{2}-2\xi_{3}^{3}\eta_{3}^{3}
-99\xi_{1}\xi_{3}\eta_{1}^{3}\eta_{2}\\[2mm]
&~~~\left.\left.+27\xi_{1}\xi_{2}\xi_{3}^{2}\eta_{1}\eta_{2}
+27\xi_{1}\xi_{3}^{2}\eta_{1}^{2}\eta_{3}
-3\xi_{1}\xi_{2}\xi_{3}^{3}\eta_{3}
\right]\right\}.\\[3mm]

\end{array}$$

\section{Conclusions}
\label{sec:con}

The rotational invariants $I_{j,k,\ell}({\bf r}_{1},
{\bf r}_{2},{\bf r}_{3})$ constructed by three spherical harmonic polynomials are the homogeneous polynomials
of orders $j$, $k$, and $\ell$ with respect to the three coordinate vectors ${\bf r}_{1}$, ${\bf r}_{2}$, and ${\bf r}_{3}$, respectively. We have rewritten the definitions for the invariants given by Biedenharn and Louck more explicitly, derived the recursive relations for the coefficients $A_{abc}$ and $B_{abc}$ using the Laplace's equations, defined two key functions $G_{a,b}$ and $F_{a,b}$ in (\ref{e-g}) and (\ref{o-f}), and calculated analytically the expressions for $A_{abc}$ and $B_{abc}$ by mathematical induction, as given in (\ref{e-2}-\ref{e-4}) and (\ref{o-2}-\ref{o-4}). Therefore, we have completely solved the problem raised by Biedenharn and Louck (p.308 of \cite{Bie}). The present method can in principle be generalized to the rotational invariants constructed by four or more spherical harmonic polynomials although the definition for the invariants depends on the order of coupling.

\begin{acknowledgments}
ZQM would like to thank Professor Fu-Chun Zhang for his warm hospitality during his visit at the University of Hong Kong where part of this work was completed. This work was partly supported by the Natural Science Foundation of China under grant No. 11174099 and No.11075014. ZCY was supported by NSERC of Canada.
\end{acknowledgments}

%\newpage
\vspace{3mm} \noindent {\bf \Large Appendix: Invariants expressed by spherical variables}

\vspace{3mm}
Choosing a special rotation such that ${\bf r}_{1}$ is along the $z$ axis and ${\bf r}_{2}$ is in the xz plane with $x$ positive, we have
$$\begin{array}{l}
{\cal Y}^{j}_{\mu}({\bf r}_{1})={\cal Y}^{j}_{\mu}(\xi_{1}^{1/2},0,0
)=\sqrt{\displaystyle
\frac{(2j+1)\xi_{1}^{j}}{4\pi}}\delta_{\mu0},\\[3mm]
{\cal Y}^{k}_{\nu}({\bf r}_{2})={\cal
Y}^{k}_{\nu}(\xi_{2}^{1/2},\theta_{12},0)\\[2mm]
~~~=(-1)^{(\nu+|\nu|)/2}\left[\displaystyle
\frac{(2k+1)\xi_{2}^{k}(k-|\nu|)!(k+|\nu|)!}{4\pi
4^{|\nu|}(k!)^{2}}\right]^{1/2}\\[3mm]
~~~~~\times (\sin\theta_{12})^{|\nu|}
P^{|\nu|,|\nu|}_{k-|\nu|}(\cos\theta_{12}),\\[3mm]
{\cal Y}^{\ell}_{-\nu}({\bf r}_{3})={\cal
Y}^{\ell}_{-\nu}(\xi_{3}^{1/2},\theta_{13},\varphi)\\[2mm]
~~~=(-1)^{(-\nu+|\nu|)/2}\left[\displaystyle
\frac{(2\ell+1)\xi_{3}^{\ell}(\ell-|\nu|)!(\ell+|\nu|)!}{4\pi
4^{|\nu|}(\ell !)^{2}}\right]^{1/2}\\[3mm]
~~~~~\times (\sin\theta_{13})^{|\nu|}
P^{|\nu|,|\nu|}_{\ell-|\nu|}(\cos\theta_{13})e^{-{\rm i}\nu\varphi},\end{array}  $$

\noindent where $P_{n}^{(\alpha,\beta)}$ is the Jacobi's polynomial (see 8.960 in \cite{Grad})
$$\begin{array}{rl}
P_{n}^{(\alpha,\beta)}(x)&=\displaystyle
\frac{1}{2^{n}}\displaystyle \sum_{m=0}^{n} \left(\begin{array}{c}n+\alpha\\m\end{array}\right)
\left(\begin{array}{c}n+\beta\\n-m\end{array}\right)\\[3mm]
&~~~\times  (x-1)^{n-m}(x+1)^{m}, \end{array}  $$

\noindent and the angles satisfy
$$\begin{array}{ll}
\cos\theta_{ab}=\displaystyle \sum_{c=1}^{3}\epsilon_{abc}\eta_{c}\left(\xi_{a}\xi_{b}
\right)^{-1/2},\\[2mm]
\cos \varphi=\displaystyle
\frac{\cos\theta_{23}-\cos\theta_{12} \cos\theta_{13}}
{\sin\theta_{12} \sin\theta_{13}},\\[3mm]
\zeta=(\xi_{1}\xi_{2}\xi_{3})^{1/2}
\sin\theta_{12}\sin\theta_{13}\sin\varphi.
\end{array} $$
\noindent Thus,
$$\begin{array}{l}
I_{j,k,\ell}({\bf r}_{1},{\bf r}_{2},{\bf r}_{3})
 =\displaystyle
\frac{\sqrt{\xi_{1}^{j}\xi_{2}^{k}\xi_{3}^{\ell}}}{k!\ell
!}\displaystyle \sum_{\nu=-k}^{k}
(-1)^{\nu}\left(\begin{array}{ccc}j & k & \ell\\ 0 & \nu &
-\nu\end{array} \right)\\[3mm]
~~~\times \displaystyle \frac{\sqrt{(k-\nu)!(k+\nu)!(\ell-\nu)!
(\ell+\nu)!}} {4^{|\nu|}}\left(\sin\theta_{12}
\sin\theta_{13}\right)^{|\nu|} \\[3mm]
~~~\times P^{|\nu|,|\nu|}_{k-|\nu|}
(\cos\theta_{12}) P^{|\nu|,|\nu|}_{\ell-|\nu|}(\cos\theta_{13})
e^{-{\rm i}\nu\varphi}.
\end{array} \eqno ({\rm A}.1) $$

From the identity
$$\begin{array}{l}
\cos\left(\nu\varphi\right)+{\rm
i}\sin\left(\nu\varphi\right)=e^{{\rm
i}\nu\varphi}=\left[\cos\varphi+{\rm
i}\sin\varphi\right]^{\nu}\\[3mm]
~~~=\displaystyle \sum_{r=0}^{\nu}
\left(\begin{array}{c}\nu\\
r\end{array}\right)\left(\cos\varphi\right)^{\nu-r} \left({\rm
i}\sin\varphi\right)^{r},\\[3mm]
\end{array}  $$

\noindent one obtains for $\nu\geq 0$
$$\begin{array}{rl}
\cos\nu \varphi &=\displaystyle \sum_{r=0}^{[\nu/2]}\sum_{s=0}^{r}(-1)^{r+s}
\left(\begin{array}{c}\nu\\ 2r\end{array}\right)\left(\begin{array}{c}r\\ s\end{array}\right)
\left(\cos\varphi\right)^{\nu-2r+2s},\\[3mm]

\sin\nu \varphi&=\sin\varphi\displaystyle \sum_{r=0}^{[(\nu-1)/2]}\sum_{s=0}^{r}
(-1)^{r+s}\left(\begin{array}{c}\nu\\ 2r+1\end{array}\right)\\[3mm]
&~~~\times \left(\begin{array}{c}r\\ s\end{array}\right)\left(\cos\varphi\right)^{\nu-1-2r+2s}.
 \end{array} $$

\noindent Substituting them into (A.1), we obtain
$$\begin{array}{l}
I_{j,k,\ell}({\bf r}_{1},{\bf r}_{2},{\bf r}_{3})
 =\displaystyle
\frac{\sqrt{\xi_{1}^{j}\xi_{2}^{k}\xi_{3}^{\ell}}}
{k!\ell!}\displaystyle \sum_{\nu=-k}^{k}
(-1)^{\nu}\left(\begin{array}{ccc}j & k & \ell\\ 0 & \nu &
-\nu\end{array} \right)\\[3mm]
~~~\times \displaystyle \frac{\sqrt{(k-\nu)!(k+\nu)!(\ell-\nu)!
(\ell+\nu)!}} {4^{|\nu|}}P^{|\nu|,|\nu|}_{k-|\nu|} (\cos\theta_{12})\\[3mm]
~~~\times P^{|\nu|,|\nu|}_{\ell-|\nu|}(\cos\theta_{13}) \displaystyle
\sum_{r=0}^{[|\nu|/2]}\sum_{s=0}^{r}(-1)^{r+s}
\left(\begin{array}{c}|\nu|\\ 2r\end{array}\right)
\left(\begin{array}{c}r\\ s\end{array}\right)\\[3mm]
~~~\times \left(\cos\theta_{23}-\cos\theta_{12}
\cos\theta_{13}\right)^{|\nu|-2r+2s}\\[3mm]
~~~\times \left[\left(1-\cos^{2}\theta_{12}\right) \left(1-\cos^{2}\theta_{13}\right)\right]^{r-s},\end{array} \eqno ({\rm A}.2) $$

\noindent for even $j+k+\ell$, and
$$\begin{array}{l}
I_{j,k,\ell}({\bf r}_{1},{\bf r}_{2},{\bf r}_{3}) =-{\rm i}\zeta\displaystyle
\frac{\sqrt{\xi_{1}^{j-1}\xi_{2}^{k-1}\xi_{3}^{\ell-1}}}
{!\ell!}\displaystyle \sum_{\nu=-k}^{k}
(-1)^{\nu}sign(\nu)\\[3mm]
~~~\times \left(\begin{array}{ccc}j & k & \ell\\ 0 & \nu &-\nu\end{array} \right)\displaystyle \frac{\sqrt{(k-\nu)!(k+\nu)!(\ell-\nu)!
(\ell+\nu)!}} {4^{|\nu|}} \\[3mm]
~~~\times P^{|\nu|,|\nu|}_{k-|\nu|} (\cos\theta_{12})P^{|\nu|,|\nu|}_{\ell-|\nu|}(\cos\theta_{13})\\[3mm]
~~~\times \displaystyle \sum_{r=0}^{[(|\nu|-1)/2]}\sum_{s=0}^{r}
(-1)^{r+s}\left(\begin{array}{c}|\nu|\\ 2r+1\end{array}\right)
\left(\begin{array}{c}r\\ s\end{array}\right)\\[3mm]
~~~\times \left(\cos\theta_{23}-\cos\theta_{12}
\cos\theta_{13}\right)^{|\nu|-1-2r+2s}\\[3mm]
~~~\times \left[\left(1-\cos^{2}\theta_{12}\right) \left(1-\cos^{2}\theta_{13}\right)\right]^{r-s}.
\end{array} \eqno ({\rm A}.3) $$

\noindent for odd $j+k+\ell$.

\end{document}